\documentclass[12pt]{article}
\usepackage{graphicx}
\usepackage{epstopdf}
\usepackage{amsmath}
\usepackage{amsfonts}
\usepackage{amssymb}
\usepackage{enumerate}
\usepackage{amsthm}
\usepackage{array}
\usepackage{epsfig}
\usepackage[small,bf]{caption}
\usepackage{appendix}
\usepackage{dsfont}
\usepackage{hyperref}
\def\1ad{\mbox{\normalsize $^1$}}
\def\2ad{\mbox{\normalsize $^2$}}
\def\3ad{\mbox{\normalsize $^3$}}
\def\4ad{\mbox{\normalsize $^4$}}
\def\5ad{\mbox{\normalsize $^5$}}
\def\6ad{\mbox{\normalsize $^6$}}
\def\7ad{\mbox{\normalsize $^7$}}
\def\8ad{\mbox{\normalsize $^8$}}

\def\beq{\begin{equation}}                     %
\def\eeq{\end{equation}}                       %
\def\bea{\begin{eqnarray}}                     
\def\eea{\end{eqnarray}}                       
\def\dj{\hbox{d\kern-0.347em \vrule width 0.3em height 1.252ex depth
-1.21ex \kern 0.051em}}

\def\half{{1\over 2}\,}

\def\ket{\rangle}
\def\bra{\langle}

\def\pt{\partial}

\def\shalf{{\mbox{$\half$}}}

\def\Dirac{\,\raise.15ex\hbox{/}\mkern-13.5mu D}
\def\dirac{\,\raise.15ex\hbox{/}\kern-.57em \partial}
\def\pslash{\,\raise.15ex\hbox{/}\kern-.57em p}

\begin{document}

                     %

\newcommand{\sheptitle}
{Ideal gas matching for thermal Galilean holography}
\newcommand{\shepauthora}
{{\sc
 Jos\'e L.~F.~Barb\'on and Carlos A. Fuertes}}

\newcommand{\shepaddressa}
{\sl
Instituto de F\'{\i}sica Te\'orica  IFT UAM/CSIC \\
 Facultad de Ciencias C-XVI \\
C.U. Cantoblanco, E-28049 Madrid, Spain\\
{\tt jose.barbon@uam.es}, {\tt carlos.fuertes@uam.es} }

\newcommand{\shepabstract}
{ 
\noindent

We exhibit a nonrelativistic  ideal gas with a Kaluza--Klein tower of species, featuring a singular behavior of thermodynamic functions at zero chemical potential. In this way, we provide a qualitative match to the thermodynamics of recently found black holes in backgrounds with asymptotic nonrelativistic conformal symmetry.}

\begin{titlepage}
\begin{flushright}
{IFT UAM/CSIC-09-19\\
}

\end{flushright}
\vspace{0.5in}
\vspace{0.5in}
\begin{center}
{\large{\bf \sheptitle}}
\bigskip\bigskip \\ \shepauthora \\ \mbox{} \\ {\it \shepaddressa} \\
\vspace{0.2in}

{\bf Abstract} \bigskip \end{center} \setcounter{page}{0}
 \shepabstract
\vspace{2in}
\begin{flushleft}
\today
\end{flushleft}


\end{titlepage}

\newpage


\setcounter{equation}{0}

\section{Introduction}
\noindent

An interesting generalization of the AdS/CFT program \cite{classic} to the case of nonrelativistic conformal field theories \cite{nrcft} was proposed in \cite{son,mcg}. The basic idea is the embedding of the Schr\"odinger group \cite{oldsch}  in $d$ spatial dimensions, ${\rm Sch}(d)$,   into
the relativistic conformal group in $d+2$ spacetime dimensions ${\rm SO}(2, d+2)$. Under this embedding,  the Schr\"odinger group arises  from a fixed light-like momentum projection $P^+ = N M$, where $N$ is a positive integer interpreted as particle number, and $M$ is the nonrelativistic mass.

 Geometrically, this projection can be achieved by a formal  compactification of the $d+2$ dimensional  conformal field theory on a light-like circle of radius $1/M$, and a similar light-like compactification of the
 ${\rm AdS}_{d+3}$ dual of the parent CFT, plus a deformation, yields the required nonrelativistic bulk description.   This results in the following family of metrics
 \beq\label{sont} ds^2 = -2\gamma^2 {r^{2z} \over
   R^{2z}} dt^2 + {r^2 \over R^2} (-2dt\,d\xi+ d{\vec x}^{\,2} ) + {R^2 \over r^2} dr^2 \;, \eeq 
   where $z$ is a real parameter controlling the relative scaling dimensions of time and spatial coordinates.  The strict case of Schr\"odinger symmetry is $z=2$, although other values of $z$ also yield interesting systems with Galilean scale invariance. The light-like coordinate $\xi$ is compactified on  a circle of size $1/M$ and the real parameter $\gamma^2$ is related to the chemical potential of the $U(1)$ isometry along the light-like circle once the system is put at finite temperature. In particular, black hole solutions with asymptotic metric given by  the $z=2$ case of (\ref{sont})  have been constructed and shown to have
   a peculiar thermodynamical scaling (cf. \cite{hotones, maldam, kov, yamada}).\footnote{See also
   \cite{more} for more on Schr\"odinger black holes.} The entropy reads
 \beq\label{peculent}
 S(T, \mu)  \propto N_{\rm eff} \, (MT)^{\,d/2} \,\left({T\over |\mu|}\right)^{d+2 \over 2}
 \;,
\eeq
 with a distinct singularity in the limit of vanishing ratio $|\mu| /T$.  In writing (\ref{peculent}) we have restored the dimensional mass parameter $M$ and the effective number of degrees of freedom $N_{\rm eff} \sim R^{d+1} /G_{d+3}$, coming from the overall power of the inverse Newton's constant in the effective Euclidean action. Other black hole metrics corresponding to hot spacetimes with $z=1$ asymptotics  in (\ref{sont}) have been constructed as Penrose limits of AdS--Kerr black holes \cite{maldam, schve}, yielding a high-temperature limit  with  the form
 \beq\label{peculentK}
 S(T, \mu) \propto N_{\rm eff} \, \left({T\over \Omega}\right)^{\,d} \, \left({T \over |\mu|}\right)\;,
 \eeq
where again we have restored the harmonic trapping frequency $\Omega$, implicit in the Penrose limits of ref.~\cite{maldam, schve}. The singular behavior of (\ref{peculent}) and (\ref{peculentK}) at small chemical potential   is quite puzzling and certainly begs for an explanation.

In this note we find an  ideal gas model which precisely matches the laws (\ref{peculent}) and (\ref{peculentK}), by simply
considering a nonrelativistic Kaluza--Klein tower of charged  `species'. These species arise naturally
from a light-like compactification of a {\it relativistic} theory with $N_{\rm eff}$ degrees of freedom in $d+2$
dimensions.  
Furthermore, we consider thermal ensembles on the metrics (\ref{sont}), which can be considered as Hawking radiation corrections to the thermodynamics of black holes, and find consistent scaling laws once the appropriate UV/IR rules are taken into account.

\section{The nonrelativistic Kaluza--Klein  gas}

\noindent

Let us consider an ideal gas of nonrelativistic particles with $N_{\rm eff}$ degenerate 
 internal degrees of freedom,
and a conserved charge
\beq\label{consch}
N = \sum_{n>0} n \,N_n\;,
\eeq
where $n$ is a further species index and $N_n$ is the  particle number of type $n$. Single particles of type $n$ contribute $n$ units to the conserved charge $N$, and have mass $M_n = n M$. 

This system can be obtained from a light-like compactification (DLCQ) of a $d+2$ dimensional system of free massless relativistic particles with $N_{\rm eff}$ degrees of freedom, i.e. we consider these fields on the metric
\beq\label{llc}
ds^2= -2\,dt\,d\xi + d{\vec x}^{\;2} \;,
\eeq
with $\xi \equiv \xi + 2\pi /M$. The conserved charge $N$ is defined as the quantized momentum in the $\xi$ direction, and $n$ appears as the Fourier index in the decomposition 
$$
\phi^a (\xi, t, {\vec x}\,) \propto \sum_{n>0}  \phi^a_n (t, {\vec x}\,) \,e^{-in\xi/M}  + {\rm h.c.}
\;,$$
where  $a$ is the internal index running over $N_{\rm eff}$ values. Notice that we have removed the zero mode $n=0$, which produces notorious problems in DLCQ. In the following, we will refer to each $\phi^a_n$
field labeled by the index $n$ as the $n$-th KK (Kaluza--Klein) species. 

\subsection{Ideal gas in a box}

\noindent

In a reflecting box of volume $V_d = L^d$, each KK species has a single-particle energy spectrum $E_n ({\vec p}\,) =
{\vec p}^{\;2} / 2M_n$, with ${\vec p}  \in {2\pi \over L} {\mathds{ Z}}^d$. The grand canonical free energy, $F(\beta, \mu)$, or rather its dimensionless version $I(\beta,\mu) = \beta F(\beta, \mu)$,  takes the form
\beq\label{freen}
I(\beta, \mu)= \beta F(\beta, \mu) = N_{\rm eff} \, \,\sum_{n> 0} \sum_{{\vec p}}  \,(-1)^F \,\log \left[1- (-1)^F e^{-\beta E_n ({\vec p}\,) + \beta \mu n }\right] \;,
\eeq
where $\beta = 1/T$ and we have allowed  for the case of a fermionic tower with $(-1)^F = -1$.  

The density  $\bra N \ket /V_d = \rho$ is then given by 
\beq\label{density}
\rho = \sum_{n>0} n \,\rho_n = \sum_{n>0} n\,{\bra N_n \ket \over  V_d} = -{1\over \beta} {\pt I \over \pt \mu}\;,
\eeq
where  the density per KK species is
\beq\label{densn}
\rho_n = {N_{\rm eff} \over V_d} \sum_{\vec p} { 1\over \exp\left[\beta(E_n ({\vec p}\,) - \mu n)\right] - (-1)^F}\;.
\eeq

The  partial thermodynamical functions at fixed $n$ are defined for $\mu\leq 0$ in the bosonic case. In the fermionic case, the fixed-$n$ functions are defined for any real $\mu$, but the complete thermodynamical functions, such as the overall density $\rho$ in (\ref{density}) will diverge for $\mu>0$ 
 upon summation in $n$. For this
reason, we shall  restrict attention to the $\mu\leq 0$ region in both Fermi and Bose cases.  

In the large volume limit at {\it fixed} $\beta \mu <0$ we can replace the momentum sums by integrals through the standard prescription
\beq\label{stint}
\sum_{\vec p} \rightarrow V_d \int{d^d p \over (2\pi)^d}
\;,\eeq
to obtain the well known result for the partial free energy density, 
$f_n(\beta,\mu)$:
\beq\label{partialfen}
\beta\,f_n (\beta, \mu) = (-1)^F \,N_{\rm eff}\,\left({M_n T \over 2\pi}\right)^{d/2} \int{d^d {\vec \ell} \over (2\pi)^{d/2}} \log\left[1-(-1)^F \exp(-\shalf {\vec \ell}^{\;2} + \beta \mu n) \right]\;.
\eeq
On evaluating the sum over the KK species, we can distinguish two regimes. At $\beta \mu \ll -1$ we
have the classical dilute gas limit, a Boltzmann gas with the KK tower dominated by the $n=1$ term with exponential accuracy:
\beq\label{bolt}
\beta f(\beta,\mu) \approx - N_{\rm eff} \left({MT \over 2\pi}\right)^{d/2} \,e^{-|\mu|/T}\;.
 \eeq
In the opposite regime, $\beta |\mu| \ll 1$, we may approximate the sum over $n$ by an integral, obtaining
\beq\label{intbetamu}
\beta f(\beta, \mu) \approx N_{\rm eff} \,(-1)^F \,\left({MT \over 2\pi}\right)^{d/2} \left({T \over |\mu|}\right)^{d+2 \over 2} \, \int{d^d {\vec \ell} \over (2\pi)^{d/2}} \int_0^\infty dx\, \log\left[1-(-1)^F e^{-x - {\vec \ell}^{\;2}/2 } \right]\;.
\eeq
We can further evaluate the integral by expanding the logarithm and integrating term by term to obtain
the final result at $\beta |\mu| \ll1$, in complete agreement with (\ref{peculent}), up to numerical coefficients:
\beq\label{finalsch}
\beta f(\beta, \mu)_\pm \approx - N_{\rm eff}\,C_\pm\, \left({MT \over 2\pi}\right)^{d\over 2} \left({T \over |\mu|}\right)^{d+2 \over 2}\;,
\eeq
where $\pm = (-1)^F$ indicates the Bose or Fermi statistics and 
\beq\label{constantinfront}
C_\pm= \pm G_{\frac{d}{2}+2} (\pm 1) \,,
\eeq
in terms of the function
\beq\label{gfunction}
G_\alpha (z) \equiv \sum_{k=1}^\infty {z^k \over k^\alpha}\;,
\eeq
with special values at $z=\pm 1$ given by Riemann's zeta function: $G_\alpha (1) = \zeta(\alpha)$, and
$G_\alpha (-1) = \zeta(\alpha) (2^{1-\alpha} -1)$.  Thus, we find essentially the same result for Bose and Fermi KK towers, up to the global factor $(1- 2^{-1-d/2})$. 

These results are obtained under the assumption that the infinite volume limit is taken {\it before} the
$\beta \mu \rightarrow 0$ limit. In the particular case of   Bose gases, the phenomenon of Bose condensation  takes place when the $V_d \rightarrow \infty$ limit is taken in conjunction with the $\beta \mu \rightarrow 0$ limit, in such a way
that the particle density on the ground state is kept fixed.

Applying (\ref{stint}) to the partial densities we obtain 
\beq\label{intap}
\rho_n = N_{\rm eff} \left({M_n T \over 2\pi}\right)^{d/2} G_{d/2} \left(z^n\right)\;,
\eeq
where we have introduced the fugacity $z= \exp(\beta \mu)$. This expression has a finite limit
as $z\rightarrow 1$, so that the temperature cannot be lowered below a critical value if the density is kept fixed. For temperatures below the critical one
\beq\label{critb}
T_c^{(n)} = {2\pi \over M_n} \left({\rho_n \over \zeta(d/2)}\right)^{2\over d}\;,
\eeq
formula (\ref{intap}) applies only to the density in {\it excited} states, $\rho_n^{(e)}$, where 
$\rho_n = \rho_n^{(0)} + \rho_n^{(e)}$, with $\rho_n^{(0)}$ representing the ground state particle density of the $n$-th
species. Hence, we have
 $$
 \rho_n^{(e)} =  \rho_n - \rho_n^{(0)} = \rho_n \left({T\over T_c^{(n)}}\right)^{d\over 2}
 $$
 in the condensation regime,  $T< T_c^{(n)}$, provided we can keep each partial density, $\rho_n$, fixed as an independent control parameter. Coming back to the grand-canonical formalism, this would entail introducing independent chemical potentials for each KK species. 

Keeping a single chemical potential, dual to the total charge $N= \sum_n n N_n$, one finds for the analog of (\ref{intap}) 
\beq\label{intapt}
\rho= N_{\rm eff} \left({M T \over 2\pi}\right)^{d/2} \sum_{n=1}^\infty n^{d/2}\, G_{d/2} \left(z^n\right)\;.
\eeq
Since the right hand side of this expression diverges as $z\rightarrow 1$, there is no lower bound on
the temperature as we eliminate $\beta\mu$ in favor of $\rho$, and correspondingly there is no critical temperature analog to (\ref{critb}). A more explicit understanding is obtained if we regularize the KK tower to be finite, $n\leq n_\Lambda$. Denoting $\Sigma (n_\Lambda) = \sum_{n=1}^{n_\Lambda} n^{d/2}$, we
have now Bose condensation below the critical temperature
\beq\label{regcrit}
T_c^{(\Lambda)} = {2\pi \over M} \left({\rho \over \Sigma(n_\Lambda) \zeta(d/2)}\right)^{2\over d}\;.
\eeq
Since $\Sigma(n_\Lambda) \sim (n_\Lambda)^{1 + d/2}$ at large $n_\Lambda$, the critical temperature
approaches zero as the KK tower regulator is removed, $n_\Lambda \rightarrow \infty$. In hindsight, we can now understand the absence of a finite critical temperature for Bose condensation as a consequence
of the spectrum having no gap at finite volume, since the nonrelativistic mass of the KK species grows without limit as $n\rightarrow \infty$.

\subsection{Ideal gas in a harmonic trap}

\noindent

The ideal gas made of KK species can also be analyzed on a harmonic trap, rather than a sharp containment box. In this case, we have a single-particle Hamiltonian
\beq\label{signph}
H_n = -{1\over 2M_n} {\vec \pt}^{\;2} + \half M_n \Omega^2 |{\vec x}|^{\;2}
\eeq
for the $n$-th species. The single-particle spectrum is controlled by the trapping frequency, independently of the value of $n$: 
\beq\label{spspec}
E_n ({\vec n}\,) = \Omega\sum_{i=1}^d n_i\;,
\eeq
where we have conventionally redefined the origin of energies to zero, by subtracting the ground state energy $E_{\rm
  vac} = (-1)^F \Omega d/2$. The resulting  grand-canonical free energy reads
\beq\label{gctrap}
\beta F(\beta,\mu) = N_{\rm eff}\,(-1)^F \sum_{n=1}^\infty \sum_{{\vec n}\in {\mathds Z}^d} \log\left[1-(-1)^F e^{-\beta \Omega \sum_i n_i  -\beta |\mu| n} \right]
\;.
\eeq
From this expression it is already clear that $|\mu|$ plays the role of an effective trapping frequency in a extra dimension. We can thus distinguish a number of regimes depending on the relative value of
the ratios $\beta \Omega$, $\beta |\mu|$, and $ \Omega /|\mu|$.

For $\beta \mu \ll -1$ we have the dilute classical gas, dominated by a single species $n=1$, up to 
corrections of relative order $\exp(-|\mu|/T)$. Hence, in this regime we obtain the standard results for an ideal gas in a harmonic trap.

For $\beta |\mu| \ll 1$ the whole tower contributes evenly, and we can approximate the sum over $n$
by an integral. In the very high-temperature limit,  $T\gg \Omega, |\mu|$, we can approximate the ${\vec n}$ sum by an integral as well, and we obtain
\beq\label{trapf}
I(\beta, \mu) \approx - N_{\rm eff}\,D_\pm \, \left({T\over \Omega}\right)^d \,\left({T \over |\mu|}\right)\;,
\eeq
where $\pm \sim (-1)^F$ refers again to the statistics and 
\beq\label{ctrap}
D_\pm = \pm G_{d+2} (\pm 1)\;.
\eeq
More explicitly, $G_{d+2} = \zeta(d+2)$, and $-G_{d+2} (-1) = \zeta(d+2) (1- 1/2^{d+1})$. 
We see that the result (\ref{peculentK}) is also reproduced in all detail, up to the value of the
numerical coefficients $D_\pm$. 

In the case that the temperature is below the gap, $T\ll \Omega$, but still $T \gg |\mu|$, we have effectively a one-dimensional system, with 
\beq\label{effone}
I(\beta,\mu) \approx -N_{\rm eff}\,2^{-F} \zeta(2)\, \left({T \over |\mu|}\right) \,\exp(-\Omega/T)\;,
\eeq
where $F=0$ for Bose statistics and $F=1$ for Fermi statistics, as usual.

\section{Hawking radiation corrections}

\noindent 

In this section we consider the thermal ensemble of radiation with $O(1)$ degrees of freedom, propagating on   the metrics (\ref{sont}). By analogy with the similar situation in relativistic AdS/CFT examples, this
would represent a piece of the $1/N_{\rm eff}$ corrections to the thermodynamic functions of the CFT (cf. \cite{brab}).  

Following the analysis of ref.~\cite{us} (see also \cite{today}), such radiation degrees of freedom have a single-particle Hamiltonian with a factorized `center of mass' dynamics with effective mass $M_n = nM$, and an `internal'
holographic dynamics given by a particular case of conformal quantum mechanics \cite{daff}. For the $z=1$ metric,
\beq\label{zunohamil}
H_n^{(z=1)}= -{1\over 2Mn} {\vec \pt}^{\;2} +  \gamma^2 M n +{1\over 2M n}\left(-{d\over d\rho^2} + {b \over \rho^2}\right)
\;.
\eeq
where $b= (d+1)(d+3)/4$ when the bulk relativistic mass of the radiation is taken to vanish. The additive shift $\gamma^2 M n$ amounts to a corresponding shift of the chemical potential $\mu \rightarrow \mu+ \gamma^2 M$. This suggests that turning on the deformation proportional to $\gamma^2$ in (\ref{sont})
is equivalent to switching on a chemical potential, a point already made in the analysis of black hole metrics in \cite{hotones, maldam}. Hence, we shall define the chemical potential after this shift is effectively subtracted, and consider a radiation free energy
\beq\label{radf}
\beta F(\beta,\mu)^{\rm rad} = I(\beta, \mu)^{\rm rad} \sim \sum_{n\geq 1} \sum_{\vec p} \log\left[1-\exp\left(-\beta (E'_n  + |\mu| n)\right)\right]\;,
\eeq
where $E'_n = E_n - \gamma^2 Mn$. The sharpest statement can be made for the case of a harmonically trapped gas, for which we add a potential
$$
V_{\rm trap} = \half M n \Omega^2 |{\vec x}\,|^2 + \half M n \Omega^2 \rho^2\;,
$$
leading to the single-particle spectrum
\beq\label{spesp}
E'_n = \Omega \left({d \over 2} +  \sum_{i=1}^d n_i + 2q + 1+\nu\right) \;,
\eeq
with $\nu^2 = b + 1/4$. We see that the radial quantum number, $q\in \mathds{Z}^+$, amounts to an extra dimension in the large temperature limit, $T\gg \Omega$. Hence, we would find a high-$T$ asymptotics
\beq\label{higtrad}
I(\beta, \mu)^{\rm rad} \sim - \left({T \over \Omega}\right)^{d+1} \left({T \over |\mu|}\right)\;.
\eeq
This result is analogous to a  well-known situation in relativistic AdS/CFT \cite{classic}, where one finds that
the entropy of gravitons in ${\rm AdS}_{d+1}$ in global coordinates scales like $T^{d}$ at high temperature, thus revealing the full dimensionality of the bulk \cite{oog}. This free energy, as well as (\ref{higtrad}), is subleading
to the black-hole free energy, proportional to $N_{\rm eff}$, in the limit that $N_{\rm eff}$ is taken to infinity faster than any other dimensionless quantities. There is a Hawking--Page transition (cf.~\cite{hp}) whenever the black-hole free energy changes sign as a function of $T$. Careful examination of
the thermodynamical functions in \cite{maldam, schve} reveals that this happens at temperatures of order $T_c \sim \Omega$. Hence, for $T< T_c$ we expect the thermodynamics to be dominated by radiation of $O(1)$ degrees of freedom.

Analogous phenomena take place in the $z=2$ case,  with single-particle Hamiltonian
\beq\label{zdoshamil}
H_n^{(z=2)} =  -{1\over 2Mn} {\vec \pt}^{\;2} +{1\over 2M n}\left(-{d\over d\rho^2} + {b_n \over \rho^2} \right)
\;.
\eeq
In this case, there is no $\gamma^2$-dependent shift, but $b_n$ depends on $n$. In particular, the
single-particle spectrum on a trap is  just like (\ref{spesp}) except for the fact that $\nu_n$ has now $n$-dependence:
$$
\nu_n^2 = \nu^2_{n=0} + 2\gamma^2 (MR)^2 n^2\;.
$$
Hence, there is an asymptotic term at large $n$ which scales like a shifted chemical potential. Performing an analogous subtraction as in the $z=1$ case ensures that the singularity is a pole at $\mu=0$ as before. 
The results are then similar to the $z=1$ case, even if required  trapped
black-hole metrics are not (yet) available in the literature for comparison. 

For untrapped metrics, such as (\ref{sont}), the radiation entropy suffers from an infrared problem as a consequence of the continuous spectrum in the radial direction. 
For heuristic purposes, we can deal with this situation by imposing a hard cutoff
in the holographic variable dictated by the UV/IR relation.  If we impose a radial cutoff $\rho \leq \rho_L$
in a system with $1/\rho^2$ potential, the spectrum will present a gap
$$
E_{\rm gap} \sim {1\over M_n \rho_L^2} 
\;.$$
If we want to associate this gap to the presence of a box of size $L$, we must  set $\rho_L \sim L$, so that
the $\rho$ variable directly represents a length scale in the boundary theory. Now, the presence of the black hole is associated to an energy scale of the order of the temperature $T$, and this corresponds to
a radial cutoff (horizon location in $\rho$ coordinates)
$$
\rho_T \sim (M_n T)^{-1/2}\;.
$$
Therefore, the effect of this cutoff on the high-energy spectrum enters in the partition function through terms like
$$
\exp\left(-\beta {4\pi^2 n_\rho^2 \over 2M_n \rho_T^2} \right) \sim  \exp\left(-2\pi^2 n_\rho^2\right)
\;,
$$
a  small $T$-independent contribution, and thus negligible in the large-$T$ limit. Hence, these considerations show that the entropy of the Hawking radiation sitting far from the horizon of a Schr\"odinger black hole
scales just like (\ref{peculent}), with the replacement of $N_{\rm eff}$ by some $O(1)$ numerical coefficient.
By the same token, this entropy in Hawking radiation may be interpreted as part of the $1/N_{\rm eff}$
corrections to the thermodynamics of the nonrelativistic CFT.

\section{Conclusions}

\noindent

The thermodynamic functions of black holes with asymptotic Galilean isometries, (\ref{peculent}), (\ref{peculentK}),  were found to diverge in the limit of large ratio $T/|\mu|$. This behavior is somewhat puzzling, especially when compared to what is expected for cold atoms at unitarity \cite{fermions, revs}, the putative physical systems to which these holographic fluids should apply (cf. \cite{kov} for a discussion). 

In this note we propose a simple explanation of the singular behavior of (\ref{peculent}) and (\ref{peculentK}), in terms of an ideal gas with a nonrelativistic Kaluza--Klein tower of species. For the cases with a concrete string theory dual, our results can be interpreted as the small $\lambda$ limit of the thermodynamic functions, with $\lambda$ the 't Hooft coupling of the higher-dimensional Yang--Mills theory and $N_{\rm eff} =N^2$, the gluon degrees of freedom. Furthermore, the situation is similar to that of relativistic AdS/CFT examples, in that only numerical coefficients would distinguish the ideal gas regime, $\lambda \ll 1$, from the black hole regime, $\lambda \gg 1$. 

 Our findings show that the nonrelativistic black hole spacetimes  considered so far, actually count degrees of freedom in full  
 $d+2$ spacetime dimensions.  The technical device of light-like compactification was used as a formal tool to achieve Galilean invariance, with local physics in the extra light-like  direction to be treated as an `artifact' of the regularization. Instead, we have seen that high-temperature black hole metrics fully excite
 local degrees of freedom in the extra dimension. 
  It is likely that any black hole spacetime whose horizon is `smeared' in the extra light-like direction will
 show this problem. Hence, the true superfluid phase of nonrelativistic systems in  $d$ spatial dimensions
 is likely to be related to black holes that localize in the extra dimension.

 In the application to systems of atoms in the unitarity limit, the conserved charge $N$ should measure
 the number of atoms. Hence, one expects the $n$-th KK species field to be related to the $n$-th power
 of the microscopic atomic field, $\phi_n \sim (\psi )^n$. This means that bound states with an arbitrary number of atoms would survive in the `classical limit' implicit in the leading gravity approximation of the bulk. This is akin to large-$N$ limits of gauge theories, where arbitrarily large `string operators' are
 kept in the leading $N\rightarrow \infty$ limit (a fact mirrored in the bulk AdS description by the infinite towers of string and Kaluza--Klein excitations kept in the leading approximation at zero string coupling $g_s \sim 1/N \sim 0$).  According to this logic, thermodynamical divergences associated to the large `length' of composite atomic operators, $(\psi)^n$, would be analogous to the divergences associated to infinitely long gauge strings, i.e. the Hagedorn critical behavior. 
 
 We therefore conclude that more work is needed in order to evaluate the applicability of current nonrelativistic AdS/CFT  constructions to explicit nonrelativistic critical points found in Nature.

\subsection*{Acknowledgements}

\vspace{0.2cm}

\noindent

This work was partially supported by the Spanish Research Ministry, MEC and FEDER
under grant FPA2006-05485,  Spanish Consolider-Ingenio 2010 
Programme CPAN (CSD 2007-00042), Comunidad Aut\'onoma de Madrid, CAM under grant HEPHACOS P-ESP-00346,  and  the European Union Marie Curie RTN network under
contract MRTN-CT-2004-005104. C.A.F. enjoys a FPU fellowship from MEC under grant AP2005-0134.



\end{document}